\pdfoutput=1
\documentclass[aps,prd,amsmath,floats,floatfix, twocolumn,
superscriptaddress,nofootinbib,showpacs,longbibliography]{revtex4-1}

\usepackage[T1]{fontenc}
\usepackage[utf8]{inputenc}
\usepackage{lmodern}
\usepackage{subcaption}
\usepackage{color,soul}

\usepackage{verbatim}
\usepackage{float}

\usepackage[dvipsnames, usenames]{xcolor}
\definecolor{linkcolor}{rgb}{0.0,0.3,0.5}
\usepackage[hypertexnames=false, unicode, colorlinks=true, linkcolor=linkcolor,
citecolor=linkcolor, filecolor=linkcolor,urlcolor=linkcolor,
pdfusetitle]{hyperref}

\usepackage[all]{hypcap}
\usepackage{graphicx}
\usepackage{xspace}
\usepackage{amssymb}
\usepackage{amsmath}
\usepackage[normalem]{ulem} 
\usepackage{bm} 

\usepackage{microtype}

\usepackage[english]{babel}
\usepackage{blindtext}

\usepackage{tikz}
\usetikzlibrary{shapes, arrows}

 \usepackage{siunitx}
\graphicspath{%
  {figs/}%
}

\DeclareMathAlphabet{\mathpzc}{OT1}{pzc}{m}{it}

\begin{document}

\title{Deep TOV to characterize Neutron Stars}

\author{Praveer Tiwari}
\email{praveer.tiwari@iitb.ac.in}
\affiliation{Indian Institute of Technology Bombay, Mumbai, Maharashtra, 400076, India}

\author{Archana Pai}
\email{archanap@iitb.ac.in}
\affiliation{Indian Institute of Technology Bombay, Mumbai, Maharashtra, 400076, India}

\hypersetup{pdfauthor={Tiwari et al.}}

\date{\today}

\begin{abstract}
Astrophysical observations, theoretical models, and terrestrial experiments probe different regions of neutron star (NS) interior. Therefore, it is essential to consistently combine the information from these sources. This analysis requires multiple evaluations of Tolman Oppenheimer Volkoff equations which can become computationally expensive with a large number of observations. Further, multi-messenger astronomy requires rapid NS characterization via gravitational waves for efficient electromagnetic follow-up. In this work, we develop a novel neural network-based map from the EoS curve to the mass and radius of cold non-rotating NS. We estimate a speed-up of an order of magnitude when compared with the state-of-the-art RePrimAnd solver and an average error of 1e-3 when calculating the mass and radius of the neutron star. Additionally, we also develop neural network solvers for obtaining EoS curves from a physics conforming EoS model, FRZ$\chi_{1.5}$. We utilize this efficient continuous map to measure the sensitivity of model parameters of FRZ$\chi_{1.5}$ towards mass and radius. We show that 8 out of 18 parameters of this model are sensitive by at least three orders of magnitude higher than the remaining 10 parameters. This information will be useful in further speeding up, as well as probing the crucial parameter space, in the parameter estimation from astrophysical observations using this physics-conforming EoS model.
\end{abstract}

\maketitle

\section{Introduction}
\label{sec:introduction}

Advancements in modern observational astronomy and theoretical high-energy simulations gave a big boost to dense matter research. Being the remnants of stellar evolution, Neutron stars (NS) are perfect laboratories for studying particles in dense environments. Observations from heavy radio pulsars~\cite{Fonseca:2021wxt, Romani:2022jhd}, the multimessenger observation of GW170817~\cite{LIGOScientific:2017vwq, LIGOScientific:2018hze}, the NICER observations~\cite{Miller:2021qha, Miller:2019cac, Riley:2021pdl, Riley:2019yda} of two pulsars, terrestrial efforts like heavy ion collision experiments~\cite{Akiba:2015jwa, Meehan:2016qon, HADES:2022gdr, Senger:2021dot} and chiral effective field theory-based calculations~\cite{Hebeler:2015hla, Lynn:2019rdt, Drischler:2021kqh, Drischler:2021kxf} have been vital in unraveling the composition of the NS. 

Future astrophysical detections are expected to put tighter bounds on the physical parameters that determine the composition of NS interiors. Particularly, joint observation of the compact binary merger with NS in GW and its electromagnetic counterpart can provide crucial information of the NS composition. This was evident with the joint observation from GW170817 and GRB170817A ~\cite{LIGOScientific:2017vwq, LIGOScientific:2018hze, LIGOScientific:2018urg}. We expect to observe more such joint coincidences in the future GW runs of advanced GW detectors. Improving the efficiency of the EM follow-up studies is important and therefore it is necessary to explore ways to speed up the detection and characterization of NS. 

NS composition is characterized by its equation of state (EoS) which is defined as the pressure as a function of the energy density in the interior. EoS along with the Tolman-Oppenheimer-Volkoff (TOV) equation gives the NS mass and radius as a function of its central pressure~\cite{Oppenheimer:1939ne}. Assuming a single EoS for every cold NS, different astrophysical observations constrain the underlying EoS model by inverting the TOV equation. These constraints help us understand the behavior of matter in extreme environments and provide input to the formation channels of the NS in the universe.

In order to invert the TOV and combine different observations, both frequentist~\cite{Huth:2021bsp, Al-Mamun:2020vzu, Fevre:2023swv} and Bayesian~\cite{Raaijmakers:2019qny, Raaijmakers:2019dks, Biswas:2020puz} approaches have been explored. In the Bayesian approach, there have been studies that assume phenomenological~\cite{Raaijmakers:2019qny, Raaijmakers:2019dks} and physics-conforming EoS models~\cite{Biswas:2020puz,Tiwari:2023tkj}. These studies assume broad priors for parameters that correspond to the dense matter near the core of the NS. This is primarily due to the inability of terrestrial experiments and theoretical calculations to simulate such an environment.

Literature exists on EoS models ~\cite{Tiwari:2023tkj, Biswas:2020puz} that efficiently combine the information from current astrophysical observations with chiral effective theory-based calculations~\cite{Drischler:2021kqh, Hebeler:2015hla, Lynn:2019rdt}. This way, we effectively probe the dense matter below the saturation density. However, these EoS models are often inherently slow to evaluate because of non-trivial functions involved in their construction. This makes the astrophysical observable-based inference even slower. The EoS models like piecewise polytropes~\cite{Read:2008iy} and spectral function~\cite{Lindblom:2010bb} are comparatively faster to evaluate. However, the inference methods will become computationally expensive to accommodate large number of NS observations that are expected in the near future, thanks to the advanced LIGO-Virgo-Kagra detectors~\cite{LIGO-aLIGODesign-NoiseCurves, VIRGO:2014yos, KAGRA_2020tym} as well as NICER instrument~\cite{Gendreau2012}. Therefore,  there is a pressing need for computationally efficient ways of characterizing neutron stars using the physics-conforming EoS models from these future astrophysical observations.

In the current machine learning (ML) era, efforts have been made to apply different ML methods to map the parameters of different EoS models to the mass and radius of the NS~\cite{ferreira2021unveiling, krastev2022translating, morawski2020neural, soma2022neural, han2021bayesian, soma2023reconstructing, fujimoto2021extensive}. Some studies invert the TOV equation to map the mass and radius of multiple NS observations to its EoS. For example, \citet{soma2023reconstructing} used two deep networks, namely the EoS solver and TOV solver, in which the TOV solver is trained to map the pressure at selected mass density inside the star to corresponding masses and radii. The EoS solver is then trained in an unsupervised manner to map these masses and radii back to the corresponding pressure points. This approach, although novel, lacks physical constraints. Therefore, it makes it less reliable to interpolate and extrapolate outside the domain of training.

In the current work, we develop a machine learning-based model to map the EoS to the radial mass and radial distance of a NS while incorporating the TOV equation during optimization. This model is a physics-informed neural network that helps ensure more reliable interpolation and extrapolation properties. The map is EoS model independent so it is compatible with both phenomenological and physics-conforming models. We also develop a framework for creating a neural network for fast evaluation of physics-conforming EoS model, namely $FRZ\chi_{1.5}$ ~\cite{Tiwari:2023tkj}
, which ensures an accurate evaluation of the speed of sound. Using this efficient map, as a first step, we study the sensitivity of the parameters of the $FRZ\chi_{1.5}$ model and show that only a select fraction of the parameters are sensitive to the observable, i.e., 8 out of 18 parameters have sensitivities higher than at least three orders of magnitude when compared with the remaining 10. This is the first time a quantified sensitivity study of the EoS model parameters has been done in the literature.

The work is organized as follows.
We discuss the details of the neural network that is developed to solve the TOV equation (deep TOV) in sec. ~\ref{sec:PINN}. The section will also detail the two EoS models, piecewise polytrope and FRZ$\chi_{1.5}$, used for training and testing the network respectively. To speed up the evaluation of FRZ$\chi_{1.5}$, we develop two neural networks which are detailed in the respective subsection. Section ~\ref{sec:results} discusses the performance of the networks developed in section ~\ref{sec:PINN}. We find that the deep TOV integrates the TOV equation approximately a hundred times faster than a typical Python ODE solver with an accuracy of more than 99\% for the radial mass and radial distance output. In the final section, sec. ~\ref{sec:conclusion}, we discuss the implications and future extension of obtaining such maps.

\section{TOV solver using Physics Informed Neural Networks}
\label{sec:PINN}

In this section, we detail the methodology of transforming the NS EoS curves to their mass and radius. We first discuss the neural network-based solver for TOV followed by the description of EoS models used for training and testing the solver.

\subsection{Deep TOV}
\label{sec:DTOV}

For a cold neutron star, the equation of state determines its mass and radius for a given central pressure through the Tolman-Oppenheimer-Volkoff (TOV) equation. It is a coupled system of ordinary differential equations that, when integrated, gives the mass and radius of the star.
The TOV equation is given by
\begin{align}
\frac{dP(r)}{dr} &= -\frac{Gm(r)\mathcal{E}(P)}{r^2}\left[\frac{\left[1+\frac{P(r)}{c^2\mathcal{E}(P)}\right]\left[ 1+\frac{4\pi r^3 P(r)}{m(r)c^2}\right]}{1-\frac{2 G m(r)}{rc^2}}\right], \label{eq:TOV1a}\\
\frac{dm(r)}{dr} &= 4\pi r^2 \mathcal{E}(P), \label{eq:TOV1b}
\end{align}
where the radial variation of pressure ($P$) and mass ($m$) inside the neutron star is expressed in terms of energy density ($\mathcal{E}$) and radial distance from the center ($r$).

We first write the TOV equation in terms of pressure as
\begin{align}
  \frac{d\tilde{m}(\tilde{P})}{d\tilde{P}} &= -
  \frac{
    2\aleph \tilde{r}(\tilde{P})^3 \tilde{\mathcal{E}}(\tilde{P})
    \left[\tilde{r}(\tilde{P}) - \tilde{m}(\tilde{P})\right]
  }{
    \left[\tilde{\mathcal{E}}(\tilde{P}) + \tilde{P}\right]
    \left[\tilde{m}(\tilde{P}) + \aleph\tilde{P}\tilde{r}^3(\tilde{P})\right]
  }, \\
  \frac{d\tilde{r}(\tilde{P})}{d\tilde{P}} &= -
    \frac
    {2\tilde{r}(\tilde{P})\left[\tilde{r}(\tilde{P}) - \tilde{m}(\tilde{P})\right]}
    {
    \left[\tilde{\mathcal{E}}(\tilde{P}) + \tilde{P}\right]
    \left[\tilde{m}(\tilde{P}) + \aleph\tilde{P}\tilde{r}(\tilde{P})^3\right]
    },
\end{align}
where,
\begin{align}  
  \aleph = \SI{\frac{4\pi R_s^3}{M_\odot c^2}}{$\text{GeV}/\text{fm}^3$},\quad
  m = \tilde{m}M_\odot, \label{eq:TOV2a}\\
  r = \tilde{r}R_s, \quad \text{where} \quad R_s = 2GM_\odot /\ c^2 \\
  P = \SI{\tilde{P}}{$\text{GeV}/\text{fm}^3$}, \quad
  \mathcal{E} = \SI{\tilde{\mathcal{E}}}{$\text{GeV}/\text{fm}^3$},\label{eq:TOV2b}   
\end{align}
and $R_s = 2GM_\odot /\ c^2$.

The solution of the TOV equations relies heavily on the EoS which appears on the right-hand side of eqn. (\ref{eq:TOV1a}) and (\ref{eq:TOV1b}) and therefore requires multiple evaluations over the course of the integration. In the absence of concrete information about the composition of NS at high densities (i.e. densities greater than the saturation density of the nucleus), we do not have a closed-form solution for this integration; making the mass and radius calculations slower. Therefore, here we use neural networks to create a computationally efficient framework for solving TOV equation.

Neural networks are a class of ML methods that mimic a neuron of a human neural system to map the input to its output. The basic unit of a neural network is a perceptron~\cite{gallant1990perceptron} which consists of a single input that maps to a hidden unit to give the output after passing through an activation function. 

Here, we use a multi-layered perceptron model and train it using a physics-informed loss function~\cite{raissi2021physics, raissi2017machine, raissi2018numerical}. The loss function of a neural network quantifies the difference between the actual and expected output of the network. For regression problems, this quantification is predominantly done through the mean squared error between the expected and actual output for discrete values. In physics-informed neural networks~\cite{raissi2021physics, raissi2017machine, raissi2018numerical}, the loss function is constructed such that the derivatives of the network are also constrained. we train the weights in such a way that the network output is reliable even outside the domain spanned by its expected output. 

In the current work, we integrate the TOV equation that gives the radial mass (m) and distance (r). Therefore, the output of the neural network is chosen to be m and r. This specification allows us to construct a loss function that imposes the TOV equation and its boundary conditions. Since, these are necessary and sufficient conditions for solving the TOV equation, the accuracy of the neural network for non-training points increases when compared with a network that is trained using a regular mean squared error loss function.

\begin{figure}
\includegraphics[scale=0.7]{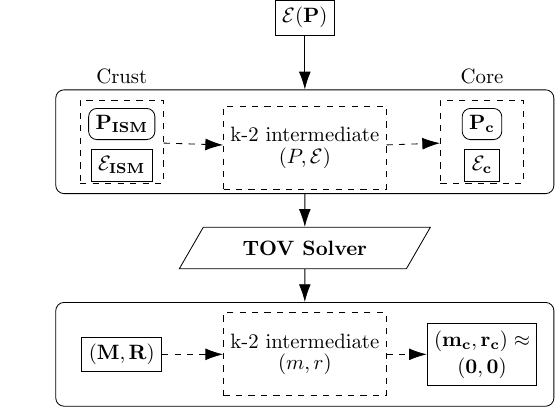}
\caption{Flow chart of the TOV solver showing the inputs (pressure and energy density) and outputs (radial mass and distance) of the network.}
\label{fig:TOVN}
\end{figure}

We divide the pressure uniformly in k steps between the central pressure and pressure at the boundary ( i.e. pressure of interstellar medium) with the points denoted by the index $i$. Then we evaluate the energy density at those points using the EoS model. These k pairs of energy density and pressure serve as the input to the network while corresponding radial masses and distances become our output (see Figure ~\ref{fig:TOVN}). The physics-informed loss function for the optimization procedure is then given by 
\begin{align}
\mathcal{L}_{\text{tot}} = \mathcal{L}_{\text{ode}} + \mathcal{L}_{\text{bc}} + \mathcal{L}_{\text{mse}} 
\end{align} 
where
\begin{align}
\mathcal{L}_{\text{mse}} &= \text{MSE}(\mathbf{\tilde{m}}, \mathbf{\hat{\tilde{m}}}) + \text{MSE}(\mathbf{\tilde{r}}, \mathbf{\hat{\tilde{r}}}), \\
\mathcal{L}_{\text{bc}} &= \text{MSE}(\mathbf{\tilde{M}}, \mathbf{\hat{\tilde{M}}}) +  \text{MSE}(\mathbf{\tilde{R}}, \mathbf{\hat{\tilde{R}}}) \\
\mathcal{L}_{\text{ode}} &= \text{L}_2 (\mathbf{\tilde{m}}, \mathbf{\hat{\tilde{m}}}) + \text{L}_2 (\mathbf{\tilde{r}}, \mathbf{\hat{\tilde{r}}}), \\ 
\text{L}_2 (x, \hat{x}) &= \sum_{i=1}^{100} \left| \frac{d\hat{x}_{i}}{d\tilde{P}_i}- \frac{dx_i}{d\tilde{P}_i}\right|^2.  
\end{align}

The hat variables are neural network output and non-hat variables are the training data. The architecture for the fully connected network used in the work is shown in Table ~\ref{tab:ATOV}. We train this network using Adam optimizer with learning rates varying from 1e-2 to 1e-6 over the course of the training.

\begin{table}[bh]
  \caption{\label{tab:ATOV}%
    \textbf{Network Architecture for TOV Solver}}
  \begin{ruledtabular}
    \begin{tabular}{ccc}
      Layer & Input-Output & Act\\
      \hline
      Input Layer  (IL) & 200-64  & tanh \\
      Hidden Layer 1 (HL1) & 64-32 & tanh \\
      Hidden Layer 2 (HL2) & 32-16 & tanh\\
      Hidden Layer 3 (HL3) & 16-32 & tanh\\
      Addition Layer 1 (AL1=HL1+HL3) & 32-32 & {}\\
      Hidden Layer 4 (HL4) & 32-64 & tanh\\
      Addition Layer 2 (AL2=IL+HL4) & 64-64 & {}\\ 
      Hidden Layer 5 (HL5) & 64-128 & tanh\\
      Output Layer (OL) & 128-200 & sigmoid
    \end{tabular}
  \end{ruledtabular}
\end{table} 
\subsection{EoS Models}

The composition of neutron stars plays a crucial role in determining their mass and radius. The outer crust, which contains heavy nuclei and free electrons, is fairly well understood and the EoS in this region is modeled through tabulated data~\cite{BPS}. As we go towards the center, the density increases, and the nuclear matter starts to drip out of the nuclei; eventually making a homogeneous mixture of mostly neutrons, a small fraction of protons, and trace quantities of electrons and other leptons~\cite{Lattimer:2021emm}. EoS models for these regions are dictated by the chiral effective field theory-based calculations~\cite{Drischler:2021kqh, Hebeler:2015hla, Lynn:2019rdt} of nuclear matter systems and heavy ion collision experiments~\cite{danielewicz2002determination, Fuchs:2005yn, Zhang:2020dvn}. Finally, the composition of the innermost regions is poorly understood and therefore the EoS models try to encompass a wide range of physics for this region~\cite{Lattimer:2021emm}.

In this subsection, we discuss three EoS models used for developing and benchmarking the Deep TOV solver detailed in the previous section.

\subsubsection{Piecewise Polytropic EoS Model}

The piecewise polytropic EoS model~\cite{Read:2008iy} gives the pressure (P) as multiple polytropes in mass density ($\rho$) stitched at different boundary points. We use this model to train the deep TOV solver by evaluating EoS curves at discretized mass density points i.e. 
\begin{align}
P(\mathcal{\rho}) =  \begin{cases} 
      K_1 \mathcal{\rho}^{\gamma_1} & \mathcal{\rho}\leq \mathcal{\rho}_1 \\
      . &  .\\
       . &  .\\
      K_n \mathcal{\rho}^{\gamma_n} & \mathcal{\rho}_{n-1}\leq\mathcal{\rho}\leq \mathcal{\rho}_n, \\ 
   \end{cases}
\end{align}
where $\gamma$s  are polytropic indices.

Here, we stitch seven polytropes to model the EoS from the crust to the core. For the crust, we use four polytropes with fixed parameters (corresponding to SLy EoS, see Table II in the appendix of ~\citet{Read:2008iy}, for values). Whereas, we use three polytropes (labelled as 1, 2, and 3) for the core. The transition densities for first to second and second to third polytropic pieces in this region are fixed to $10^{14.7}$ $g/cm^3$ (say, $\rho_1$) and $10^{15.0}$ $g/cm^3$ (say, $\rho_2$) respectively. We vary these three polytropes to obtain 26782 EoS curves. More precisely, the logarithm of the pressure (in $dyne/cm^2$) at $\rho_1$ (i.e. $\log(p_1)$) and the three polytropic indices of the core are sampled uniformly from (34.3, 34.9), (1, 4.5), (0, 8), and (0.5, 8), respectively. These are further vetted if they are not able to produce 0.8 $M_{\odot}$ NS. These curves are then used to train and test the TOV solver.  

\begin{figure*}[!htb]
\includegraphics[scale=0.5]{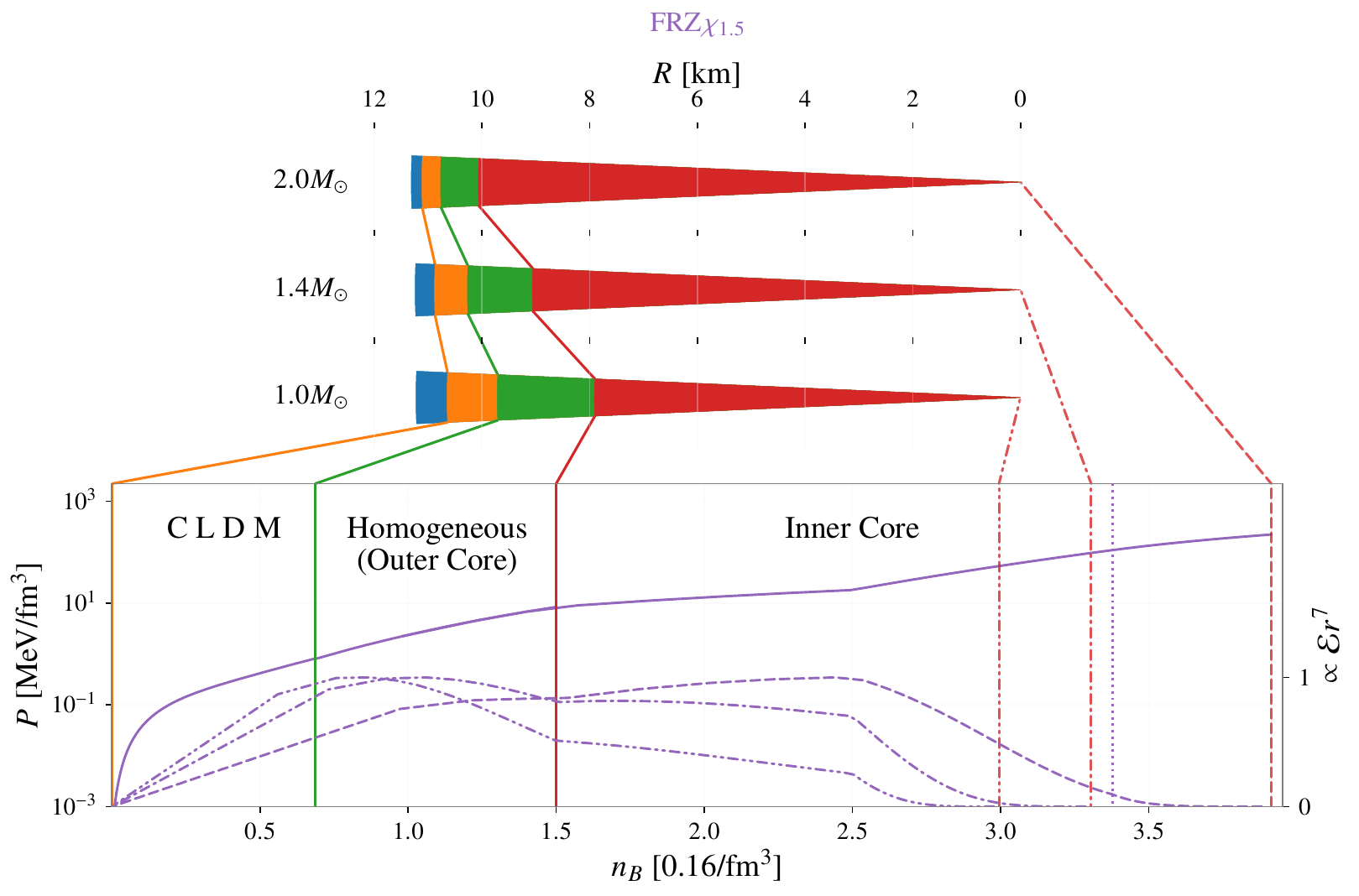}
\caption{Characteristic EoS curve for $FRZ \chi_{1.5}$. The top wedges show the radius of the three neutron stars along with the contribution coming from different portions of the stars' interiors. The solid curve corresponds to the pressure variation as a function of number density. Dashed, dot-dashed and double dot-dashed curves represent quantity which is proportional to $\mathcal{E} r^7$ (which indicates tidal deformability contribution) for $2.0M_{\odot}$, $1.4M_{\odot}$ and, $1.0M_{\odot}$ NS, respectively.}
\label{fig:PFRZ}
\end{figure*}

\subsubsection{Physics Conforming EoS Model: $FRZ\chi_{1.5}$}
\label{sec:PCEOS}

In this work, we use one example of physics conforming EoS; namely  FRZ$\chi_{1.5}$ EoS model~\cite{Tiwari:2023tkj} which  we illustrate in the Figure (\ref{fig:PFRZ}). The outer crust (denoted by a very thin blue patch in the figure) is modeled by tabulated data~\cite{BPS}. The orange patch is the inner crust which is modeled through a compressible liquid drop model~\cite{Chamel2008}. The outer core, denoted by green patch, is modeled as a homogeneous matter~\cite{Forbes:2019xaz, Tiwari:2023tkj} while for the inner core, a piecewise polytropic model has been used. The EoS model for homogeneous matter expands the energy of the nuclear matter about the proton fraction. This expansion reduces the model dependence on the symmetry energy parameters (as we will find in section \ref{sec:sensitivity}). This means that the constraints on the symmetry energy parameters from the heavy ion collision experiments~\cite{danielewicz2002determination, Fuchs:2005yn, Zhang:2020dvn} have minimal effect on this model, giving us an independent probe to the NS. Additionally, the functional form of the pure neutron matter-energy has been developed to incorporate the constraints coming from the chiral effective field theory calculations~\cite{Drischler:2021kxf}. For the inner core, in the absence of prior knowledge of its composition, the model attaches three polytropes at 1.5 times the saturation density with varying transition densities.  

Such models, although directly connecting the physical EoS parameters with observables, are relatively slow to evaluate. For example, the computation of transition density between the inner crust and the outer core involves a root finder which slows down the EoS evaluation for FRZ$\chi_{1.5}$. Other steps such as finding the coefficients of nuclear matter energy expansion are computationally expensive because of the matrix inversion step involved. We discuss next how we remedy these inefficiencies through neural networks.
\begin{figure}[ht]
\includegraphics[trim={1.05cm 0 0 0},clip, width=\linewidth]{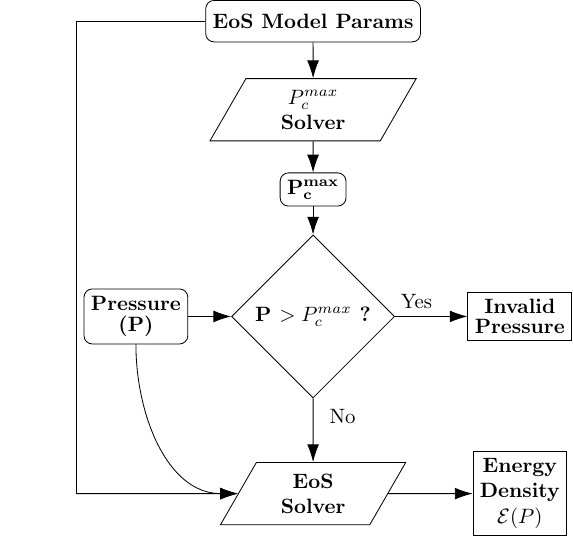}
\caption{Schematic diagram showing how the EoS model parameters and pressure are mapped to the energy density using the two neural networks, namely EoS solver and $P_c^{\text{max}}$ solver.}
\label{fig:EoSN}
\end{figure}

{\it EoS Solver for $FRZ\chi_{1.5}$}: Here, we develop a computationally efficient pressure-energy density map for a given parameter of $FRZ\chi_{1.5}$. More precisely, a neural network solver (called an EoS Solver) has been developed that evaluates the energy density for a given parameter value of $FRZ\chi_{1.5}$ and pressure. Additionally, we note that every equation of state curve has a maximum central pressure beyond which it cannot support an NS. Therefore, for a given FRZ$\chi_{1.5}$ parameter, we first evaluate the maximum central pressure allowed using a neural network, named $P_c^{\text{max}}$ solver. We compare the output (i.e. $P_c^{\text{max}}$) of this solver against the input pressure. If the pressure is less than the $P_c^{\text{max}}$, only then we proceed to the EoS solver. This solver uses the input pressure and FRZ$\chi_{1.5}$ parameters to output the energy density. (see Figure \ref{fig:EoSN})
The network architecture of the two solvers is detailed in the table \ref{tab:AEOS}. We use the same optimization procedure as the one chosen for the TOV solver (see section \ref{sec:DTOV}). The loss function for the EoS solver is a sum of two mean squared errors. The first term is the mean squared error between the output and the energy density, and the second term is between the inverse of the derivative of the network and the square of the speed of sound for the training data. The total loss function, therefore, becomes
\begin{align}
\mathcal{L}_{\text{tot}} =  \text{ MSE}(\mathcal{E}, \hat{\mathcal{E}}) + \text{ MSE}\left(c^2, \left[\frac{d\hat{\mathcal{E}}}{dP}\right]^{-1}\right)
\end{align}

\begin{table}[bh]
  \caption{\label{tab:AEOS}%
    \textbf{Network Architecture for EOS and $P_c^{max}$ Solver}}
  \begin{ruledtabular}
    \begin{tabular}{ccc}
      Layer & Input-Output & Act\\
      \hline
      Input Layer  (IL) & 19-8  & tanh \\
      Hidden Layer 1 (HL1) & 8-32 & tanh \\
      Hidden Layer 2 (HL2) & 32-16 & tanh\\
      Hidden Layer 3 (HL3) & 16-32 & tanh\\
      Addition Layer 1 (AL1=HL1+HL3) & 32-32 & {}\\
      Hidden Layer 4 (HL4) & 32-32 & tanh\\
      Hidden Layer 5 (HL5) & 32-32 & tanh\\
      Addition Layer 2 (AL2=AL1+HL5) & 32-32 & {}\\ 
      Hidden Layer 6 (HL6) & 32-64 & tanh\\
      Hidden Layer 7 (HL7) & 64-32 & tanh\\
      Addition Layer 3 (AL3=AL2+HL7) & 32-32 & {}\\
      Output Layer (OL) & 32-1 & sigmoid
    \end{tabular}
  \end{ruledtabular}
\end{table}

\subsubsection{Spectral Equation of State Model}

Here,  we discuss the spectral EoS; a thermodynamic stable EoS model that is expressed in terms of the adiabatic index, $\Gamma (P)$,  and is given by
\begin{align}
\Gamma (P) = \frac{\mathcal{E} + P}{P}\frac{dP}{d\mathcal{E}}.
\end{align}
Spectral expansion of the $\Gamma (P)$ is performed in terms of dimensionless pressure~\cite{Lindblom:2010bb}, as
\begin{align}
\Gamma(P) = \exp \left[\sum_{k=0}^{3} \gamma^{(s)}_k \left(\frac{P}{P_0}\right)^{k}\right],
\label{eq:spec}
\end{align} 
where, $P_0$ is the scale factor that is chosen to be the minimum pressure in the NS, and $\gamma^{(s)}$s are spectral polytropic indices. This expansion used with equation (\ref{eq:spec}), to get 
\begin{align}
\mathcal{E}(P) = \frac{\mathcal{E}_0}{\mu(P)} + \frac{1}{\mu(P)} \int_{P_0}^{P} \frac{\mu(P')}{\Gamma(P')} dP', \quad \text{where,} \\
\mu(P) = \exp \left[- \int_{P_0}^{P} \frac{1}{P'\Gamma(P')} dP'\right].
\end{align}

The two EoS models, that we discussed before, contain transitions between different polytropes. These transition points do not have a well-defined speed of sound. Such discontinuities affect the accuracy of the deep TOV output because our solver creates a smooth map from the EoS to the radial mass and distance. Therefore, we expect the spectral representation of EoS, having a thermodynamically stable prescription with a continuous speed of sound, will conform to the smooth TOV map constructed via the neural networks. 
Thus, for testing, the deep ToV,  we generate the EoS curves by sampling $\gamma^{(s)}_{0-3}$ uniformly from the region [0.2, 2], [-1.6, 1.7], [-0.6, 0.6], [-0.02, 0.02], respectively. Additionally, we also impose the condition that the $\Gamma$ should be in the range [0.6, 4.5].

\section{Results}
\label{sec:results}

\begin{figure*}[!htb]
\begin{subfigure}{0.48\textwidth}
\includegraphics[scale=0.8]{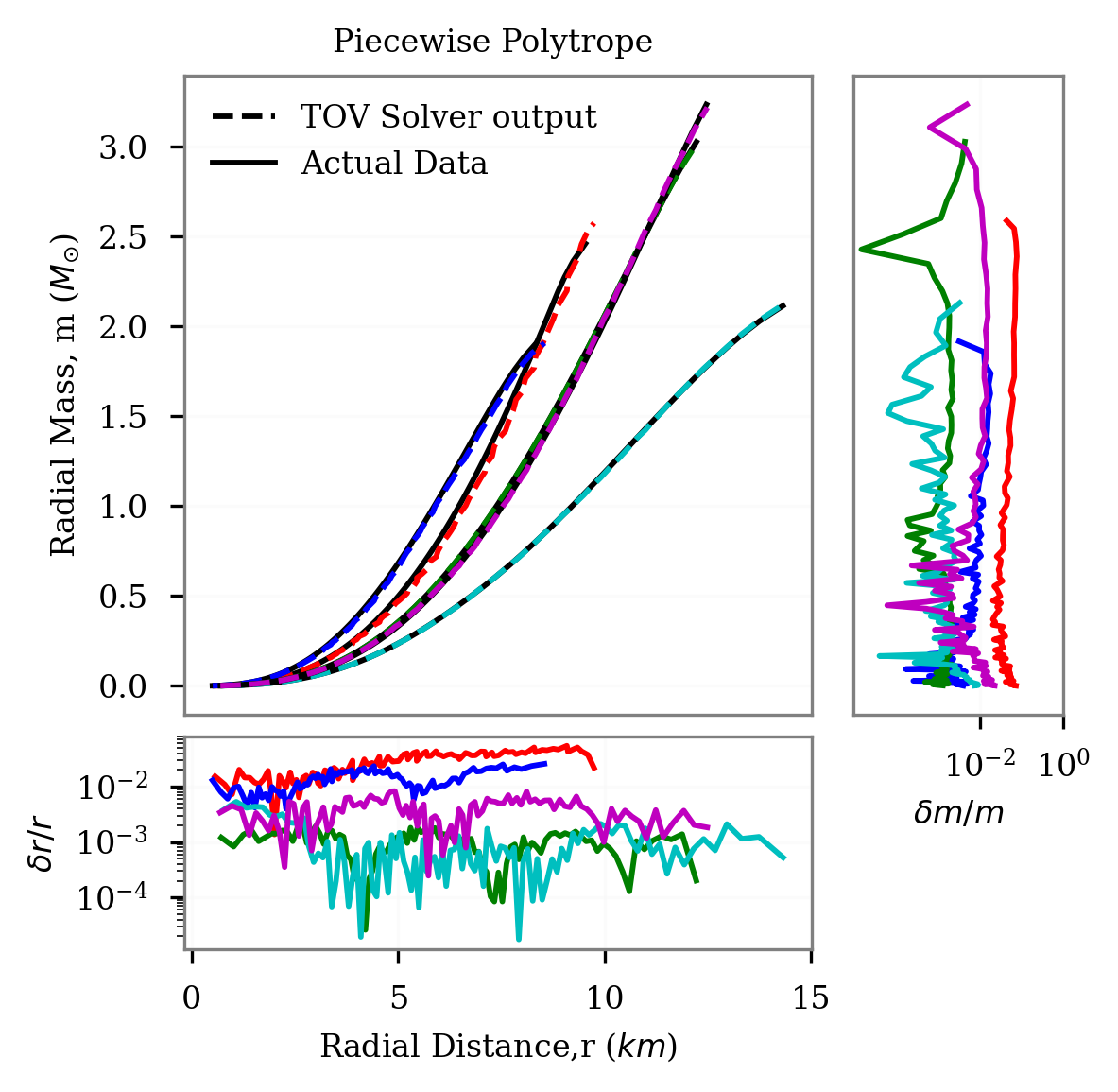}
\caption{m-r curves for validation EoS curves. Comparing the ODE solver output (solid black lines) with neural network output (dashed colored lines). The bottom and side plots show relative errors in radial distance and mass respectively.}
\label{fig:TOVvalPP}
\end{subfigure}
\begin{subfigure}{0.48\textwidth}
\includegraphics[scale=0.8]{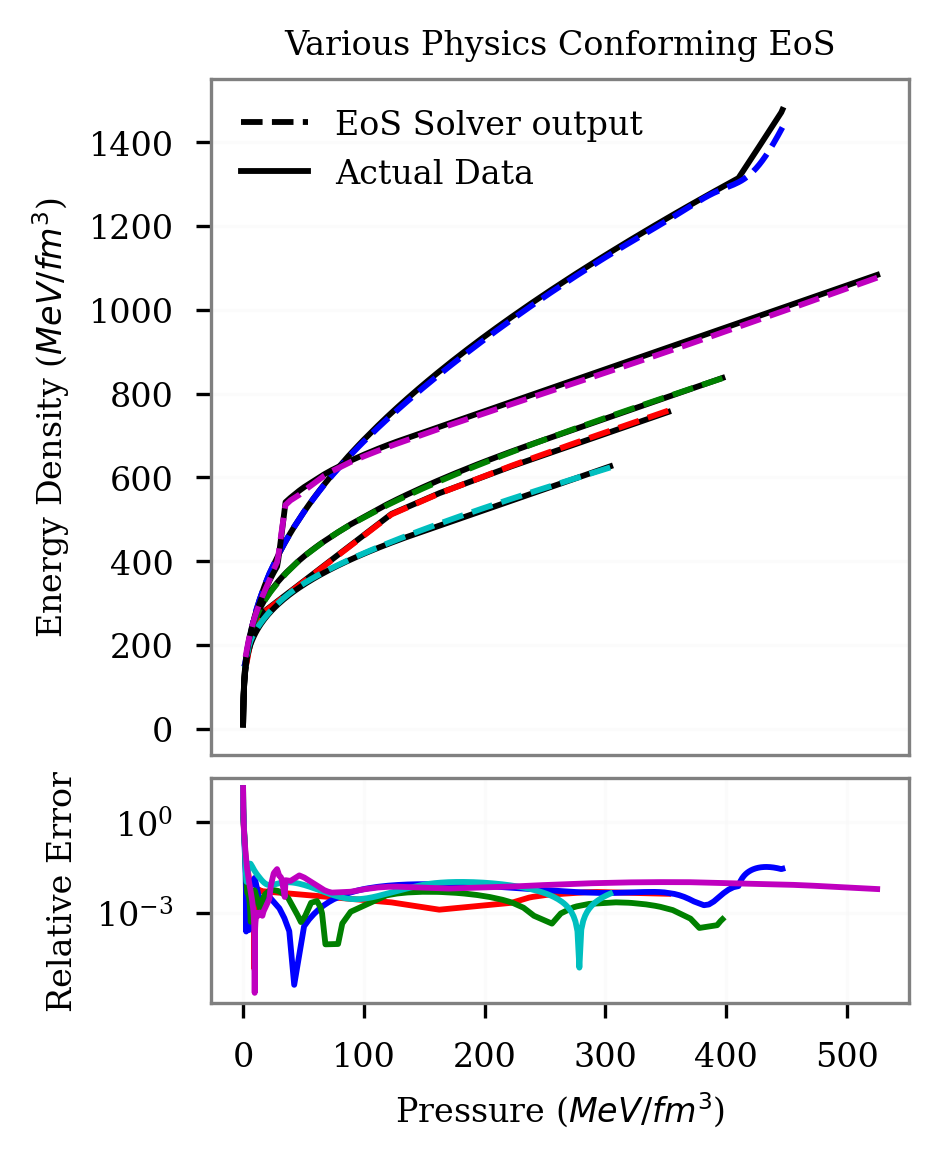}
\caption{EoS ($\mathcal{E}(P)$) for different parameters of $FRZ\chi_{1.5}$ model. Comparing the network output (dashed) with the actual EoS curve (solid) using the testing data.}
\label{fig:EoSval}
\end{subfigure}
\caption{Validation accuracy for (a) deep TOV and (b) EoS solver.}
\end{figure*}

We train the deep TOV solver using the EoS curves from the piecewise polytrope EoS model. We use both the mean squared error and physics-informed loss function, for comparison. The training time significantly increases for the latter and we find that the training errors were comparable in both cases. The network trained with the mean squared loss function performs well for validation and testing data as well. This feature can be attributed to the input and output units for our network. We train the network to map the EoS curves to the radial mass and distance which imposes the physical constraints similar to the one coming via the physics-informed loss function. Therefore, we could relax the conditioning on the loss function and train the network using the mean squared error function instead.

However, for the EoS solver, the physics-informed loss function is required to make sure that the derivative of the network can be related to the speed of sound of the EoS. Also, we use min-max transformation (using the training data) to scale the input and output of all the networks in order to train the network faster.

\subsection{Validation of Solvers}

We train the deep TOV solver using EoS curves from the piecewise polytrope (PP) EoS model. The performance of the network is visualized using the validation PP EoS curves in Figure (\ref{fig:TOVvalPP}). The deep TOV solver output is plotted along with the solution of the TOV equation using the RK-4 ODE solver. Looking at the error for the final point of the curves in the figure, we find that the relative error for computing the total mass and radius of the NS from these EoS curves is less than 0.01. This error can be further reduced by using an EoS curve with a continuous speed of sound prescription at all points, contrary to piecewise polytropes, which have discontinuities at the transition points of the polytropes. Since the deep TOV is a smooth map from the input to the output, any non-smoothness in the target function space can affect the accuracy.

In Figure (\ref{fig:EoSval}), we demonstrate the accuracy of the EoS solver to generate EoS curves from $FRZ\chi_{1.5}$. We are able to achieve an average accuracy of more than 99\% in retrieving energy density from the input pressure and $FRZ\chi_{1.5}$ parameters. The network was trained in a way so that it can preserve the speed of sound but $FRZ\chi_{1.5}$ does not have a continuous speed of sound in the inner core region. Therefore, it puts a limit on the accuracy of capturing the EoS curve. We find that we could recover the speed of sound curves in the regions of the EoS model where it is defined. The recovery, however, has oscillations near the transition points where the speed of sound is ill-defined. This is one of the results of this paper. We hope to extend this study in the future with an EoS model with a continuous speed of sound prescription.

\subsection{Testing Solvers (Robustness)} 

\begin{figure*}[!htb]
    \begin{subfigure}{0.48\textwidth}
    \includegraphics[scale=0.8]{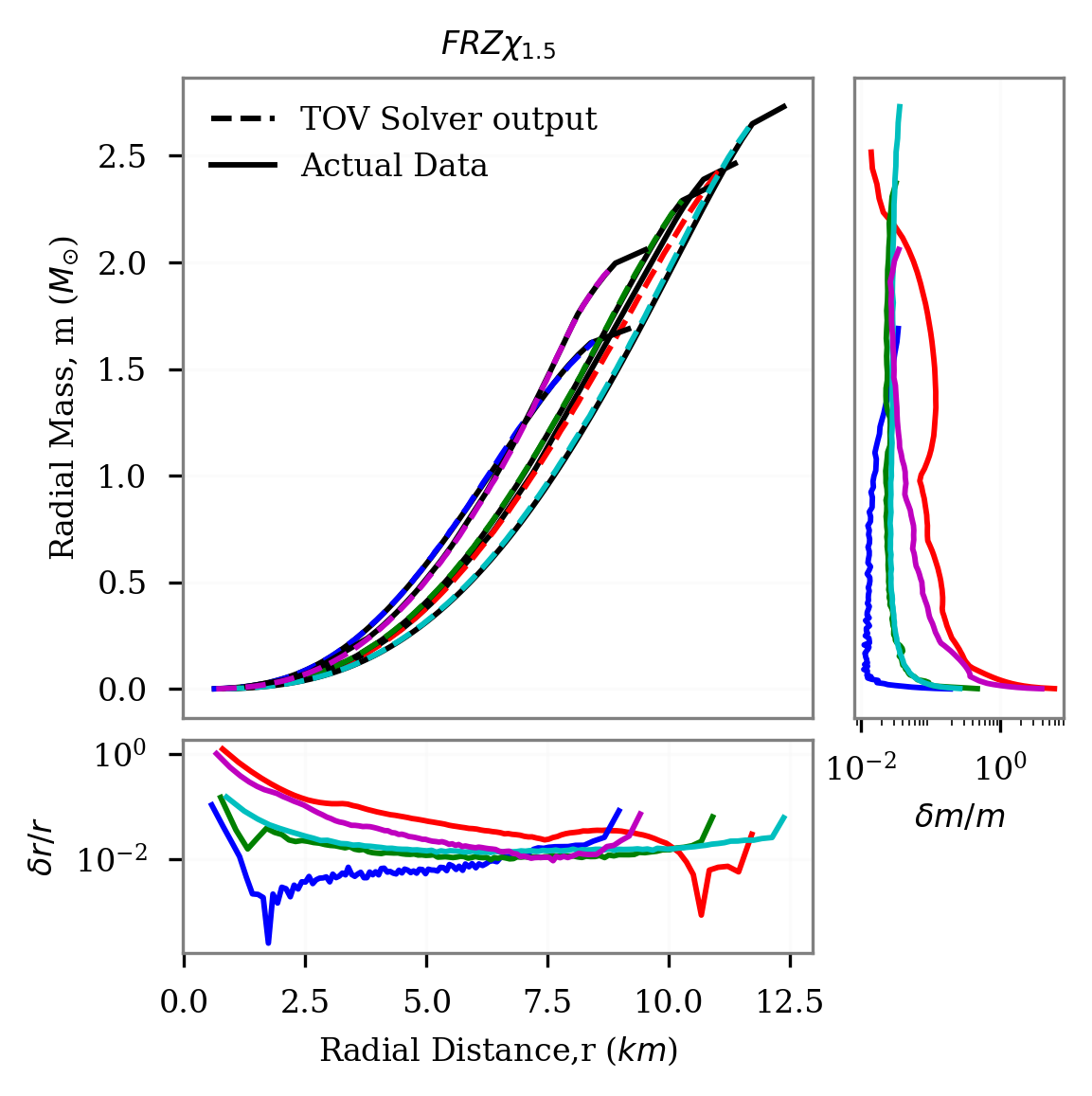}
\caption{}
     \label{fig:TOVvalFRZ}
     \end{subfigure}
    \begin{subfigure}{0.48\textwidth}
     \includegraphics[scale=0.8]{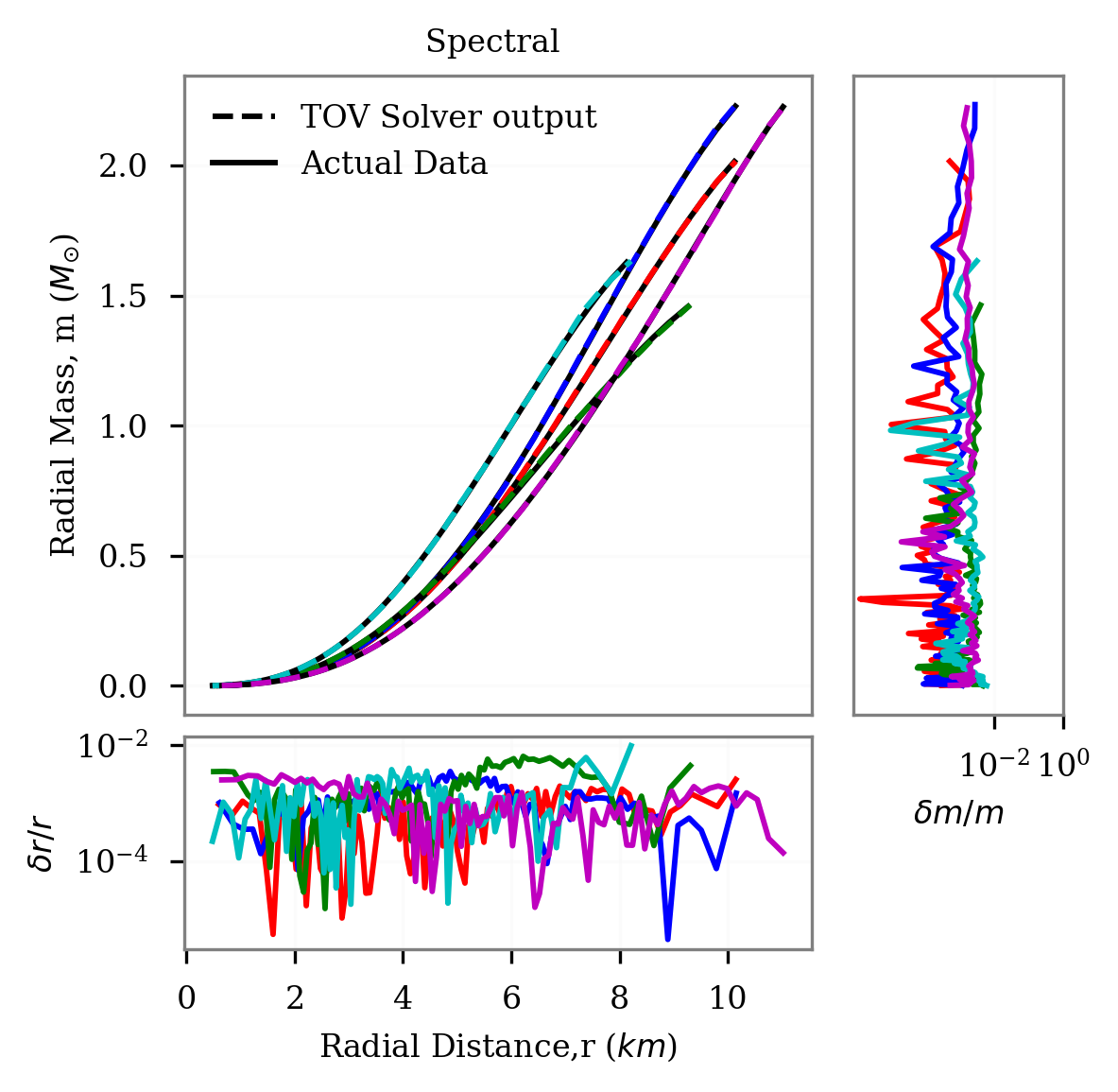}
     \caption{}
     \label{fig:TOVvalSpec}
     \end{subfigure}
     \caption{m-r curves for test EoS curves from (a) $FRZ\chi_{1.5}$ and (b) Spectral EoS model. Comparing the ODE solver output (solid black lines) with neural network output (dashed colored lines). The bottom and side plots show relative errors in radial distance and mass respectively.}
\end{figure*}

In Figure (\ref{fig:TOVvalFRZ}), we plot the output (and corresponding error) of the deep TOV solver for calculating the radial mass and the distance from the $FRZ\chi_{1.5}$ EoS curves. The accuracy of the solver again lies around 99\% mark largely due to the presence of the speed of sound discontinuity in the inner core of the NS. In principle, some contribution to the error, in this case, comes from the inaccuracy of the EoS solver. We test the network with actual $FRZ\chi_{1.5}$ curves as well and found that the error in mapping the EoS curve via EoS solver has a minimal effect on the final radius and mass estimate for $FRZ\chi_{1.5}$ EoS curves. 

We test the deep TOV solver using the EoS curves from the spectral EoS model. The solver performs well owing to the smoothness of the spectral EoS curves. The average error in both the radius and the mass of the NS is found to be around 0.1\% (see Figure (\ref{fig:TOVvalSpec})). We do not use spectral EoS curves for training or validation of the network and the functional form for the spectral curves is also very different from the piecewise polytropes. Therefore, an improved performance, in this case, points to the fact that the deep TOV is trained well to solve the TOV equation and it has minimal bias from the training data. Hence, we expect that the deep TOV solver will generalize well for smooth prescriptions of EoS.

\subsection{Speed Up}

We compare the speed of numerically integrating the TOV equation by lsoda~\cite{Petzold1983} with our TOV solver and we find that the neural network-based solver is faster than the lsoda~\cite{Petzold1983} method by roughly three orders of magnitude. 

We also compare the speed with the state-of-the-art, RePrimAnd solver~\cite{PhysRevD.103.023018}.  We find at least a factor of $\sim 8$ increase in the speed with comparable accuracy for the spectral EoS curves. More precisely, on a single-core CPU, we find that the RePrimAnd takes $\sim 151$ ms whereas deep TOV takes $\sim 19$ ms to evaluate masses and radii of 1000 NSs. Note that, we evaluate deep TOV solver through seven matrix multiplications between weights and inputs at each layer. During the speed benchmarking above, we performed these multiplications through regular Python operations. We can, in principle, make this matrix multiplications more efficient. Additionally, these operations can easily be performed in parallel using multiple cores. It will be interesting to see how the speed increases with the number of cores and how it compares with the RePrimAnd solver.

%
    

\begin{figure}
\includegraphics[width=\linewidth]{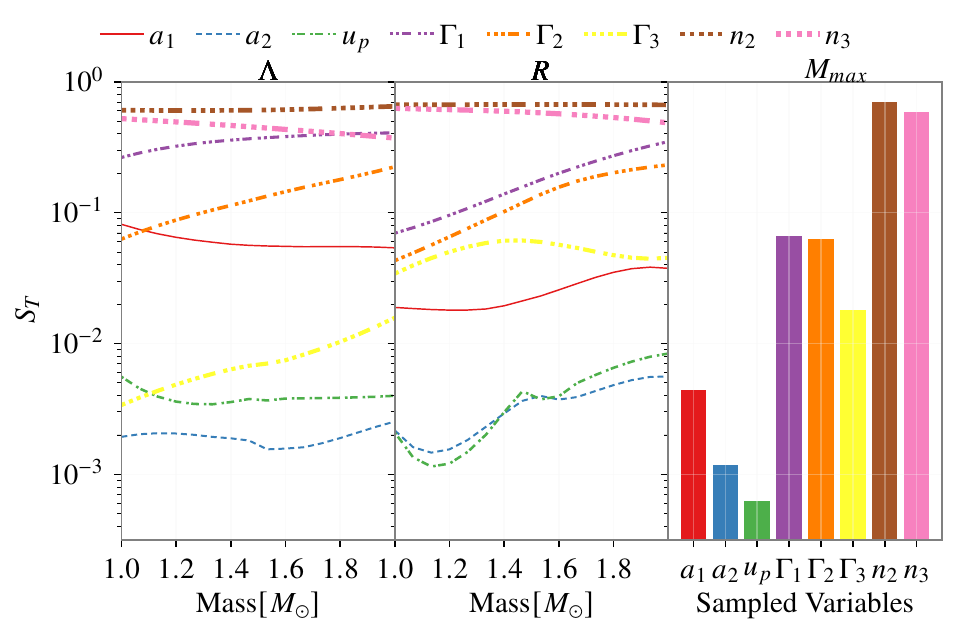}
\caption{Total effect index of top eight most sensitive EoS model parameters for $FRZ\chi_{1.5}$ towards $\Lambda$, $R$, and $M_{max}$. The remaining variables have at least an order of magnitude less $S_T$}
\label{fig:SIT}
\end{figure}

\subsection{Sensitivity Study for FRZ$\chi_{1.5}$}
\label{sec:sensitivity}
A computationally efficient mapping from the EoS model to the mass-radius observable can be used to explore the relationship between the model parameters and the observables. We start by probing the sensitivity of the model parameters toward observables. This way, we hope to understand the physical constraints that can be placed based on the observation of neutron stars. 

We employ variance-based sensitivity analysis on the FRZ$\chi_{1.5}$ model in the hopes of understanding the dependence of its parameters on observables. The quantity known as, total effect index ($S_T$) {\it aka} Sobol index ~\cite{Sobol2001}, given by 
\begin{align}
S_{T} (X_i) = \frac{E_{X_{\sim i}}\left[V_{X_i}(Y(\mathbf{X})|X_{\sim i})\right]}{V(Y(\mathbf{X}))},
\end{align}
quantifies how a particular model parameter $X_{i}$ affects the value of the observable $Y(\mathbf{X})$ directly or indirectly through other model parameters. Here, $E$ represents expectation, $V$ represents variance, and $X_{\sim i}$ represents all the model parameters except $X_i$. The expression in the denominator quantifies the total variance of the observable, $Y$, calculated from the samples. The numerator quantifies how much of that variance can be attributed to the variation in the values of the variable, $X_i$.  Therefore, the value of  $S_T$ can vary between 0 to 1 having values close to 1 for highly sensitive variables.

We perform this analysis for the $FRZ\chi_{1.5}$ to measure the sensitivity of the $FRZ\chi_{1.5}$ model parameters towards $Y=\Lambda$, $R$, and $M_{\text{max}}$.  We assess the sensitivity of $\Lambda$ parameter by integrating the asymptotic metric expansion equation using the lsoda method for cross-checking the trends in $R$ and $M_{\text{max}}$ that was obtained from the deep TOV solver and $P_c^{\text{max}}$ solver respectively. We find that out of 18 parameters required to construct $FRZ\chi_{1.5}$ model, 10 parameters have $S_T$ which is several orders of magnitude less than the highly sensitive ones for all the three observables.  As can be seen from Figure (\ref{fig:SIT}), the parameters that are responsible for determining the EoS at densities near saturation point and above, dictate the values of the observables. 
A crucial implication of the sensitivity study, and one of the main results, is that we can employ the physics-conforming EoS model in the Bayesian inference study to estimate the neutron star properties using only sensitive, informative EoS parameters from the astrophysical observations.

\section{Conclusion}
\label{sec:conclusion}

The TOV equation is used for understanding and characterizing neutron stars through simulations and observations. This work aims to speed up these analyses by creating an efficient map between EoS curves to mass, and radius observables. 

We use a novel approach with neural networks to construct the deep TOV solver for evaluating the radial mass and distance as a function of EoS curves that are specified at discrete points. Since the activation functions for the network are smooth functions, deep TOV is a smooth function and therefore we use them to map smooth EoS curves to radial mass and distance. We demonstrate this by calculating the errors in recovering observables from the spectral EoS model. We find the average error to be less than 0.1\% in getting the observables. These maps are on average 15 times faster than the state-of-the-art RePrimAnd solver~\cite{PhysRevD.103.023018}.

Further, to improve the computational efficiency of Bayesian inference using physics-conforming EoS models,  we create a neural network-based map that
computes the energy density as a function of pressure and model parameters.   We check this framework on the FRZ$\chi_{1.5}$ model. Using this EoS solver,  we investigate the sensitivity of the parameters of FRZ$\chi_{1.5}$ model using the Sobol index towards the radius and maximum mass of NS. We note that the study shows that 8 out of 18 model parameters are sensitive.  The Sobol index-based sensitivity evaluation for FRZ$\chi_{1.5}$ model parameters is another main result of this work.  It clearly shows that such sensitivity studies can help in identifying the crucial model parameters that can be used to probe the physics of NS using the observations.  For example,  we can extend this map to asymptotic metric expansion equations to evaluate tidal deformability parameters which can be probed from GW observations,  which we wish to pursue as a future study.

These efficient maps and frameworks can be useful in implementing complex EoS models to simulate the NS systems for understanding their evolution and possibly correlating it with detected astrophysical observations. This will involve expanding the EoS specification in the finite temperature limit and testing the validity of these maps for such EoS models.

Additionally,  in the future,  we wish to improve the map using deeper networks at the cost of the speed of the solver.  Further improvement in accuracy may be possible by increasing the input and output dimensions of the TOV solver to incorporate more pressure points. The TOV solver developed here is independent of the EoS used so we can use EoS solvers from different EoS models to map it to the mass, and radius observable. There are more sophisticated techniques for solving ordinary differential equations~\cite{2018arXiv180607366C, Lagaris1997ArtificialNN}. We hope to explore these methods in the future to further speed up these maps without sacrificing the accuracy and physics. With the next generation of GW detectors being right around the corner, we expect to have a meteoric rise in the number of binary neutron star detections which will make rapid estimation of EoS crucial.

\begin{acknowledgments}

We thank Wolfgang Kastaun and Rahul Kashyap for their valuable comments. AP acknowledges the support from the SERB-Power fellowship
grant SPF/2021/000036, DST, India, and SPARC, MOE grant number SPARC/2019-2020/P2926/SL. This document has LIGO DCC number LIGO-P2400158.
\end{acknowledgments}
\pagebreak
\bibliography{References}

\end{document}